\documentclass[preprint2]{aastex}
\shorttitle{Radio Exploration of Planetary Habitability}
\shortauthors{Wolszczan et al.}

\begin{document}
\title{Radio Exploration of Planetary Habitability: Conference
	Summary}

\author{%
T.~Joseph~W.~Lazio (Jet Propulsion Laboratory, California Institute of
Technology), 
A.~Wolszczan (Dept.\ Astronomy \& Astrophysics, Center for Exoplanets and Habitable Worlds, Pennsylvania State University), 
M.~G\"udel (Dept.\ Astrophysics, University of Vienna),
Rachel A.~Osten (Space Telescope Science Institute),
Jan Forbrich (Dept.\ Astrophysics, University of Vienna),
M.~M.~Jardine (School of Physics \& Astronomy, University of St.~Andrews), 
\and
P.~K.~G.~Williams (Harvard-Smithsonian Center for Astrophysics)}

\begin{abstract}
Radio Exploration of Planetary Habitability was the fifth in the
series of American Astronomical Society's Topical Conference Series.
Notable aspects of the conference included the interdisciplinary
nature of both the topics and the intellectual breadth of the
participants, the diversity of approaches to studying this topic
presented by recent discoveries and of the participants themselves,
the expanding meaning of the topic of ``star-planet interactions,''
and the expectation of an increasingly statistical approach to the
topic.  Potential areas of future research include the actual extent
to which planetary magnetic fields shield planetary atmospheres; the
planetary dynamo process itself, particularly once multiple extrasolar
planetary magnetic fields are confirmed; and ``planet-star
interactions.''  
A major major topic of the conference concerned observational
opportunities, highlighted by a number of new or upcoming, specialized observatories
to observe exoplanets especially at radio wavelengths.
This article summarizes these main points of the
conference and expands briefly upon these potential avenues for future
investigation.  A future meeting on this topic, given the variety of
data sets being generated over the next few years, is warranted.
\end{abstract}



\section{Motivation}\label{sec:motivate}

\begin{figure*}[bht]
\centering
\includegraphics[width=\textwidth]{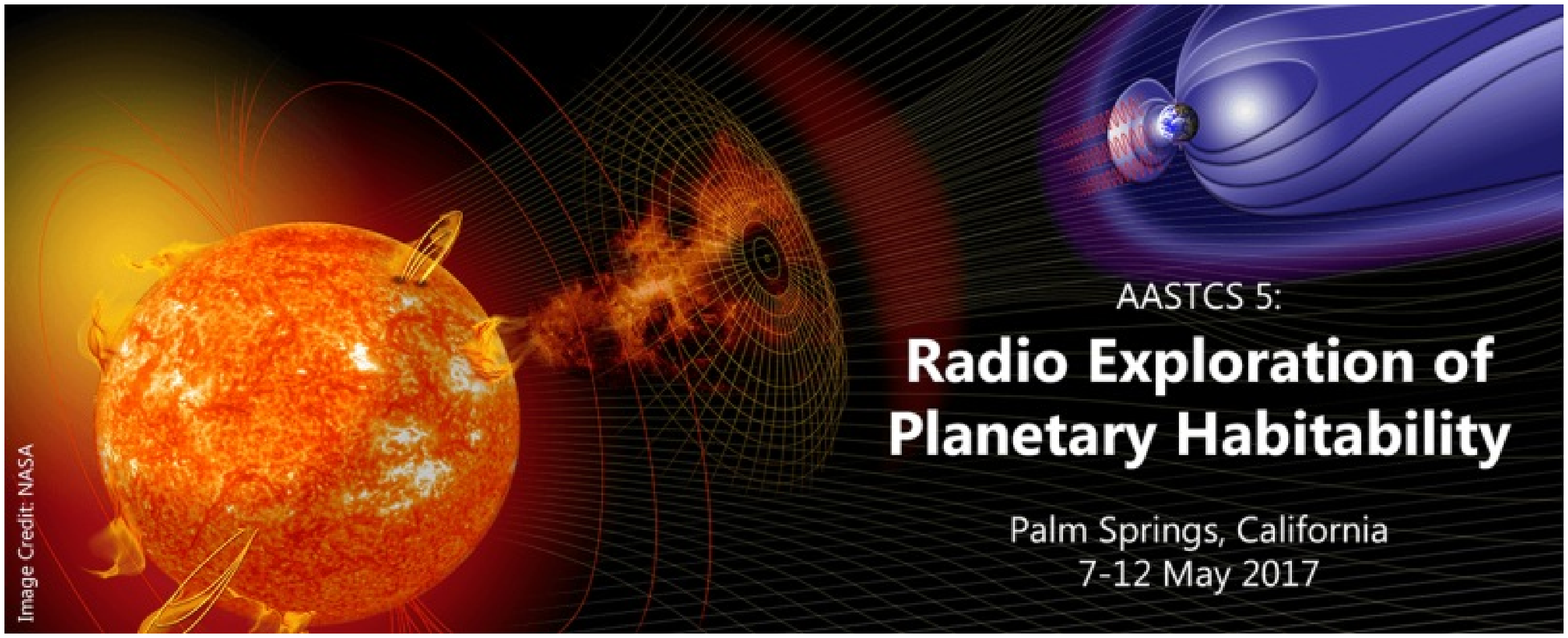}
\end{figure*}

Low-mass stars, i.e., stars with masses significantly below one solar mass, have been regarded as exciting targets for planet
searches, because of their abundance compared to the Sun-like stars,
and their desirable properties from the standpoint of sensitivity of
radial velocity and transit surveys.  On-going and upcoming
ground-based searches, such as \hbox{CARMENES} and the Habitable
Planet Finder, and space missions such as K2, Gaia, and {TESS} are
expected to deliver a large number of potentially habitable planets.
Planets around low-mass stars with biosignatures detectable over
interstellar distances are likely to be located close to their parent
stars, motivating a growing need to investigate stellar activity and
its influence on habitability of possible planetary companions.

In our solar system, recent results from the MAVEN spacecraft have
provided spectacular and persuasive evidence that it was the activity
of the young Sun that stripped Mars of its original atmosphere.  In
addition to low-mass stars, stellar activity of young, solar-type
stars may also be a consideration in assessing the current and future
habitability of planets.  Even though the period of high activity
is typically much shorter for solar-type stars than for low-mass stars; nevertheless, it does cover
the era of the formation of a crust, an ocean, and a secondary atmosphere
on a planet like Earth.

Complementing studies of stellar activity, research on planetary
habitability must also include detection of extrasolar planetary
magnetic fields across the entire range of planet masses and ages.
There are a variety of potential techniques for detecting and studying
planetary magnetic fields, but, guided by the experience in our solar
system, the radio emission generated by magnetic, kinetic, and
unipolar star-planet interactions is the most promising approach.  The unipolar interaction between an
unmagnetized object and a magnetized object, for which the Jupiter-Io
interaction is the exemplar, could generate radio emission from main
sequence stars with orbiting planets, but also from planets around
white dwarfs, planets around brown dwarfs, and moons around extrasolar
gas giants.

The recognition that radio wavelength observations represent a
powerful approach for broadening the study of planetary habitability
comes at a time when a new generation of broadband receivers and high
time-frequency resolution data acquisition hardware offer unique tools
for detailed investigations of the phenomena that are relevant to
these topics.

Of course, all the above issues cannot be addressed properly without
further theoretical and observational developments in various areas,
among which we emphasize magnetic field generation in fully convective
stars, stellar magnetic activity, and coronal mass ejection (CME)
generation as factors that are crucial for planetary habitability, as
are number of related topics.

Radio Exploration of Planetary Habitability was the fifth in the
series of American Astronomical Society's Topical Conference Series
designed to explore these topics.  The following sections summarize
broad themes resulting from the conference (\S\ref{sec:summary}) and
highlights from the sessions (\S\ref{sec:spi}
and~\S\ref{sec:observe}).

\section{Overview}\label{sec:summary}

The week was marked by a number of stimulating presentations and
vigorous discussion.  In part, this discussion was enabled by a
meeting structure in which portions of each session were devoted to
open discussion among all participants, led by session chairs.  Many
participants commented that they found this structure, and the
preservation of blocks of time for discussion, to be among the most
productive aspects of the meeting, allowing questions and issues to be
explored more fully.

In a meeting of this length, it is not possible to summarize all of
the results.  Presentations from the meeting are available at the \citet{AAS}.
Nonetheless, several themes were apparent.

Both the topic and the participants were highly interdisciplinary.
The topic necessarily draws on experience from within the solar system
and from multiple extrasolar planetary systems.  This
interdisciplinary nature presents both opportunities and challenges.
The Sun and solar system planets can be studied at much higher time
and spatial resolution than can other stars and extrasolar planets.  Participants with expertise in heliophysics and planetary science
presented multiple examples of physical processes that are witnessed
only incompletely in the atmosphere of another star or in the
interaction between a stellar wind and a planet's magnetosphere
whereas they can be
studied at high resolution in the solar system; in the best case,
spacecraft can be sent to solar system planets to provide \textit{in
situ} measurements.  Examples might include the lift-off of a CME from
the Sun, the impact of a CME on the atmosphere of an unmagnetized
planet, or the electromagnetic interactions between Jupiter and its
moon Io.  There were members of various mission science teams present,
and future meetings on this topic and related topics are encouraged to
work to include representation from relevant mission science teams.

Conversely, the opportunity of high resolution in the solar
system presented an obvious challenge, particularly in communication.
When only relatively coarse temporal or spatial resolution is
available (e.g., whole-disk averages of a star), detailed models may
provide little guidance, and inferences from our particular old and inactive
star or even extrapolations to stars of different activity levels, ages, or
masses may risk questionable conclusions.

A notable example of the benefit of the interdisciplinary nature of
this conference and a potential area for future research concerns the
protective nature of planetary magnetic fields.  Planetary
magnetic fields commonly are taken as a requirement for providing
shielding of planetary atmospheres from the solar wind, with the MAVEN
observations of a CME's erosion of the Martian atmosphere taken as a
\textit{prima facie} example.  Nonetheless, planetary scientists note
that the mass loss from the atmospheres of Venus, Earth, and Mars are
approximately equal.

Another common theme was diversity.  Extrasolar planetary systems
contain planets not present in the solar system (e.g., super-Earths)
allowing a broader range of planetary properties to be explored.
Moreover, the stellar hosts of these extrasolar planets cover a
diverse range of ages, activity levels, and spectral types, allowing a
potential exploration of planetary habitability in a manner not
accessible in the solar system.  In an extreme example, the stars
\objectname{UV~Ceti} and \objectname{BL~Ceti} are a binary and are
presumably coeval, yet show dramatically different magnetic field
structures and levels of activity.  Many participants stressed the
importance of being able to construct a more complete understanding of
the possible development or evolution of planetary habitability as a
result of the study of the various objects.

A prime example of diversity in this conference and an avenue for
future research concerns the generation of planetary magnetic fields
themselves.  In the solar system, vastly different classes of planets
appear able to generate and sustain planetary dynamos (terrestrial,
ice giant, gas giant).  There are clear differences between the
characteristics of the magnetic fields of these different classes of
planets, which is both suggestive that many extrasolar planets are
likely to generate and sustain magnetic fields but also frustrating
because the constraints on planetary dynamos are provided currently by
only a small number of planets.

Then, there is diversity in the subtleties. From afar, Venus and Earth
are sister planets, with nearly equal masses, perhaps similar
originally outgassed atmospheres (of carbon dioxide, nitrogen, and
water); but they have evolved in largely different directions, one
supporting plate tectonics, a strong magnetic dynamo, and a water
ocean while the other maintains a stagnant lid, has no dynamo and
therefore lacks a magnetosphere while it keeps a dense carbon dioxide
atmosphere devoid of water. The challenge to characterize extrasolar
planets to this level is obvious, as important as these features may
be for habitable environments.

The diversity of the participants themselves was also notable.  As
previously noted, their intellectual breadth spanned the traditional fields of
planetary science, heliophysics, and astrophysics, with many
participants knowledgeable in multiple aspects of these fields.
Geographically, participants hailed from five continents.  There was a
deliberate effort to ensure experience and gender balance, with
early- and mid-career and senior scientists all present and making
presentations; of the 42 total presentations, seventeen were invited
and women presented six of these (35\%).

The role of star-planet interactions is an area of interdisciplinary
research, a representation of the diversity of the field, and an area
for future work.  The focus of star-planet interactions is often on
the effects of the star on the planet; examples include the
aforementioned atmospheric erosion and the chemistry of a planet's
atmosphere in response to a star's illumination (and changes
thereof).  Many participants noted that these star-planet
interactions are likely to change with time, as the star evolves, and
the growing number of extrasolar planetary systems allows this process
to be studied in more depth and across the spectral and evolutionary
sequence of stars.

However, it is increasingly recognized that the \emph{planetary
effects on the star} may also be important.  Such processes may no
longer be apparent in the solar system (though they may have been
important in the early solar system), but they could include
interactions between planetary and stellar magnetic fields, leading to
changes in the stellar magnetic field, rotation, and potentially even
evolution.  A related aspect is that there is not be a single, unique
evolutionary sequence in stellar properties; for example, evolution of
a star's rotation rate may be affected by whether or not it has a
close-in planet and whether or not that planet is magnetized; further
evolutionary differences likely depend upon  the initial rotation period of the
host star  and its subsequent spin-down due to a dynamo-driven wind.

Finally, multiple participants stressed the importance of moving
toward large samples.  The \textit{Kepler} mission is perhaps the
exemplar of how large and homogeneous data sets can open new
opportunities.  Missions such as Gaia and TESS and ground-based
surveys (including, but not exclusively limited to, LSST), large data
sets and comprehensive statistical analyses should become, and are
becoming, the norm, rather than a future aspiration.

\section{Stellar Activity and Star-Planet Interactions}\label{sec:spi}

This session covered a diversity of topics, ranging from theoretical
investigations of dynamo models (both planetary and stellar) to
observational constraints that ran the gamut of stellar rotation
period measurements to pulsed radio emission.  A point emphasized by
multiple speakers is that that the study of star-planet interactions has a requirement of understanding \emph{both} the star and the planet. Many talks focused on one or the other of the system to further this understanding. The talks revealed that while much progress has been made in understanding stellar and planetary interactions, there is still much to be learned.  The discussion periods were stimulated by thought-provoking comments in the talks.

With the expansion in capabilities over the past decade, the diversity
of the magnetic properties of stars is much more apparent, with
numerous new developments in understanding summarized.  A connection
to fundamental stellar parameters does not necessarily result in the
ability to completely predict those magnetic properties: stellar twins
are not necessarily coronal twins, with the stars \objectname{UV~Ceti}
and \objectname{BL~Ceti} being prime examples.   Observational
constraints on stellar winds can be determined, but attempts to detect
stellar coronal mass ejections directly, using a variety of low radio
frequency instrumentation, has not met with success yet.

An emerging bimodality in rotation periods for low-mass stars suggests
rapid rotational evolution.  The diversity in magnetic topology may
indicate bistability in dynamo behavior, a somewhat controversial
theoretical idea. Clues from the geodynamo and solar system
observations help with the vast parameter space between planets, brown
dwarfs, and stars. Some other thought-provoking ideas: a
re-examination of the inner bound of the traditionally defined
habitable zone may be needed, to account for the impact of the stellar
magnetic topology and its creation of severe space weather. Habitable
zones for M~dwarfs also intersect the zone where star-planet
interactions occur, making for likely false positive detections of
``habitable'' planets.

\section{Observations of Stars, Brown Dwarfs, and
	Exoplanets}\label{sec:observe}

This session was focused on observations across the entire mass range
of objects.  Due to events occurring within both the \textit{Cassini}
and Juno missions, no members from either science team were able to
attend, but recent results have highlighted some of the exciting
observations from those missions as they apply to the magnetic fields
of gas giants.

Given that energetic particles easily generate radio emission by a
variety of processes (synchrotron, gyro-synchrotron, gyrotron, and
plasma resonances), and often in concert with magnetic fields, many of
the presentations focussed on radio wavelength observations.  The
diversity of radio telescopes available to execute such searches for
evidence of planetary magnetic fields or star-planet interactions was
apparent, with participants describing observations with established
telescopes (Arecibo, Green Bank Telescope[GBT], Very Large Array
[JVLA], Very Long Baseline Array [VLBA]) and new telescopes (Long
Wavelength Array-Owens Valley Radio Observatory [LWA-OVRO], the Low
Frequency Array [LOFAR], the Murchison Widefield Array [MWA], the
Hydrogen Epoch of Reionization Array [HERA]).  In many cases, the
search for radio emission from sub-stellar objects was not one of the
motivating rationales for the telescope, yet it is becoming clear that
these telescopes are well suited for such observations.

Particularly for the new telescopes, large fields of
view or significant observing programs or both are feasible, enabling
large data sets to be constructed (with either many targets monitored
or long time series produced or both).  In view of the Sun's 11~yr
activity cycle, the need for a diverse target set and long duration
observations is clear.

Radio wavelength observations as part of a multi-wavelength suite of
observations was also apparent, with participants describing observing
projects motivated by the \textit{Kepler} results and joint radio-UV
or joint radio--X-ray observations motivated by solar observations.
To date, the main recent focus for stellar high-energy work has been
X-ray observations, but new observational capabilities have led to a
renaissance of stellar radio astronomy and its complementary
perspective on high-energy phenomena. Combining both X-ray and radio
observations in the time domain, it is possible to study the entire
sequence of magnetic energy release, from particle acceleration to
energy transformation and heating. One signature of this scenario is
the so-called Neupert effect of correlated radio and X-ray light
curves, which has been discovered in solar observations and has also
been found in nearby active stars. With better radio sensitivity, it
is now becoming possible to extend such observations to different
types of active stars, yielding new constraints on the high-energy
irradiation of their vicinities, with potential consequences for
habitability.

\acknowledgements
We thank the sponsors of this conference, who helped make it possible
and particularly enabled many early-career researchers to attend.
Sponsors of the meeting were the Associated Universities, Inc., the
NASA Exoplanet Science Institute, the John Templeton Foundation, the
PennState Center for Exoplanets and Habitable Worlds, and the
Universities Space Research Association.
We thank the entire Scientific Organizing Committee for crafting an
exciting program; members of the SOC were A.~Wolszczan, 
 I.~Baraffe, T.~Bastian, E.~Berger, M.~Browning,
A.~Burgasser, P.~Driscoll, J.~Forbrich, G.~Hallinan, C.~Heiles,
Th.~Henning, J.~Kasting, J.~Lazio, J.~Linsky, R.~Osten,
K.~Poppenhaeger, A.~Reiners,  A.~Segura, D.~Werthimer, P.~K.~G.~Williams, and P.~Zarka.
Part of this research was carried out at the Jet Propulsion
Laboratory, California Institute of Technology, under a contract with
the National Aeronautics and Space Administration.

\end{document}